\documentclass{article}

\usepackage{template}
\pdfoutput=1

\usepackage{pgfplotstable,booktabs}
\usepackage[utf8]{inputenc} 
\usepackage[T1]{fontenc}    
\usepackage{hyperref}       
\usepackage{url}            
\usepackage{booktabs}       
\usepackage{amsfonts}       
\usepackage{nicefrac}       
\usepackage{microtype}      
\usepackage{lipsum}
\usepackage{graphicx}
\usepackage[normalem]{ulem}
\useunder{\uline}{\ul}{}
\usepackage{longtable}
\usepackage{colortbl}
\usepackage{tikz}
\usepackage{tabu} 
\graphicspath{ {./figures/} }
\usepackage{mathtools}
\usepackage{placeins}
\usepackage{isotope}
\usepackage{makecell}
\usepackage{lscape}
\usepackage{caption}
\usepackage{blindtext}
\usepackage{subcaption}
\usepackage{textgreek}
\usepackage{svg}
\usepackage{indentfirst}
\usepackage{amssymb}

\pgfplotstableset{
   begin table=\begin{longtable},
   end table=\end{longtable},
}

\title{Scholar Ranking 2023: Ranking of Computer Science Departments Based on Faculty Citations}

\author{
    Sai Shi \\
    Department of Computer \& Information Science \\
    Temple University \\
    Philadelphia, PA 19122 \\
    \texttt{sai.shi@temple.edu} \\
    \And
    Aniruddha Maiti \\
    Department of Computer \& Information Science \\
    Temple University \\
    Philadelphia, PA 19122 \\
    \texttt{aniruddha.maiti@temple.edu} \\
    \And
    Ashis Kumar Chanda \\
    Department of Computer \& Information Science \\
    Temple University \\
    Philadelphia, PA 19122 \\
    \texttt{ashis@temple.edu} \\
    \And
    Slobodan Vucetic \\
    Department of Computer \& Information Science \\
    Temple University \\
    Philadelphia, PA 19122 \\
    \texttt{vucetic@temple.edu} \\
}

\begin{document}
\maketitle

\begin{abstract}
Scholar Ranking 2023 is the second edition of U.S. Computer Science (CS) departments ranking based on faculty citation measures. Using Google Scholar, we gathered data about publication citations for 5,574 tenure-track faculty from 185 U.S. universities. For each faculty, we extracted their \emph{t10} index, defined as the number of citations received by their 10th highest cited paper. For each department, we calculated four quality metrics: median \emph{t10} (m10), the geometric mean of \emph{t10} (g10), and the number of well-cited faculty with \emph{t10} above 40\% (c40) and 60\% (c60) of the national average. We fitted a linear regression model using those four measures to match the 2022 U.S. News ranking scores of CS doctoral programs. The resulting model provides Scholar Ranking 2023, which can be found at \url{https://chi.temple.edu/csranking}. 

\end{abstract}

\section{Introduction}
A previous version of the Scholar ranking \cite{Vucetic2017FacultyCM} was published in the spring of 2017, based on citation data collected during the fall of 2016. This previous effort demonstrated that it is possible to learn a simple formula from citation measures that has a high correlation with peer assessment scores of CS doctoral programs published by the U.S. News (USN). 

A few years have passed since the last publication of the Scholar ranking, and a new U.S. News ranking came out in 2022\footnote{\url{https://www.usnews.com/best-graduate-schools/top-science-schools/computer-science-rankings}}. Given the fact that the data on which the last ranking was performed is a few years old, we felt that it would be helpful to conduct another round of data collection and validate our method with the recent U.S. News ranking. The first objective is to refine the data collection method and collect a new set of high-quality faculty citation data. The second objective is to use the 2022 U.S. News ranking to validate the method proposed in the first version of the scholar ranking and observe changes in the ranking. The third objective is to analyze the trends in aggregated metrics used to perform the ranking given the data sets, with the first collected during the fall of 2016 and the other during the fall of 2021.

\section{Data collection}
In this section, we explain the data collection process which took place from September 2021 to December 2021. The data collection team consisted of two CS graduate students and a CS professor.

\subsection{U.S. News (USN) Data}
USN is well-known for producing several rankings. We gathered the scores from the most recent ranking of CS doctoral programs, \emph{Best Computer Science Schools}, which was published in 2022. USN collected the names of those to be surveyed for the science doctoral surveys in the summer of 2021. We retained the scores from the 2013 version of \emph{Best Computer Science Schools} from previous data collection. 

USN ranks programs using scores generated from surveys sent to academic professionals\footnote{\url{https://www.usnews.com/education/best-graduate-schools/articles/science-schools-methodology}}. Only survey responses from fall 2021 and early 2022 were used to compute the scores. The surveys asked respondents to rate each program from 1 to 5, with one being marginal and five being outstanding. Respondents could skip programs and select "don't know" if they were unfamiliar with them. Each program's score is the average of its survey ratings if it has at least ten ratings. Programs with less than ten ratings are not scored. Unlike the scores reported in USN 2017, where the program is ranked if it has a score of at least 2.0, USN 2022 published and ranked the scores of programs that are lower than 2.0. USN does not provide raw survey data or information about potential sources of bias in responses. USN does not attempt to fill in missing values. 

\subsection{Computer Science Faculty List Data}

We collected the data on 5,574 tenure-track CS professors from 185 departments ranked by USN. We identified 2,011 faculty on our 2022 list but not on our 2017 list, including 1,750 new professors and 261 professors who joined another department. In contrast, 4,728 professors were collected in 2017 from 173 departments. The number of CS professors included in our list increased by 17.89\%.

We consider a professor part of a department if the professor is listed on the website's faculty list. We found each website by performing Google searches on the school's name followed by "computer science" or "cs." In most cases, lists of faculty and their appointments were on pages labeled "Directory," "People," or "Faculty." Some pages did not specify appointments. In these cases, we found a professor's appointment by performing a Google search on their name and exploring their website or profile page. 

We only consider tenure-track professors, which would have the rank of assistant, associate, or full professor. We excluded professors who have the titles "Clinical", "Courtesy", "Adjunct", "Research", "Teaching", "Emeritus", "Visiting", or other additional labels that indicate that the professor was not a tenure-track professor.


For universities with CS departments, we added all professors because they were in a department for CS professors only. In some universities, the computer science faculty are part of joint departments called "Electrical Engineering and Computer Sciences" or "Computer Science and Engineering." Some universities have colleges or departments of computing or informatics, which contain faculty in CS, library science, information sciences, or management information systems. These departments made it harder to distinguish who was a CS professor. We determined that professors with research interests and publications in CS topics will be CS professors. We looked at the publications or research interests on their department profile page or website. CS topics include artificial intelligence, machine learning, data science, human-computer interaction, bioinformatics, cybersecurity, and others. Some cases we do not consider within CS are sensor networks, hardware, genomics, signals, and others.

There were some unique cases in choosing the departments. New York University has the Department of Computer Science and Department of Computer Science and Engineering within two separate colleges. We only considered the department within the Courant Institute of Mathematical Sciences. Case Western University has the Department of Computer and Data Sciences and Department of Electrical, Computer and Systems Engineering. We only included professors within The Department of Computer and Data Sciences because the faculty of the other department generally had research interests that we excluded. Rochester Institute of Technology has a big college of Computing and Information Sciences, which consists of several departments, such as Computer Science, Computing and Information Sciences, Software Engineering, etc. We only included faculty from the department of Computer Science since they listed their faculty in separate departments.  

Another issue was affiliated professors and professors who have non-primary appointments. We did not include affiliated faculty if they were in a separate section from the main faculty or labeled as having a joint or secondary appointment. In cases when affiliated, joint, or secondary appointed professors were mixed in with primary faculty, we added all professors who were on the list because the department listed them as its professors.

Outside these departments, some professors have made significant contributions in CS venues but have primary appointments within engineering, biology, statistics, business, or other departments. We did not include non-CS professors because we needed a time-effective and unbiased method of finding these professors.

In 2017, we found that 23.6\% (1114/4728) of CS faculty are assistant professors, and that percentage has changed to 29.2\% (1630/5574) in 2022. Since assistant professors are starting to establish their publications, we treat them differently from associate and full professors. We refer to associate and full professors as senior faculty, and our collection of Google Scholar data focused on senior faculty.

\subsection{Google Scholar Data}
After determining what professors are in each department, we identified each professor's Google Scholar page to collect their citation measures\footnote{\url{https://scholar.google.com/}}. Despite some limitations of automated web
crawling \cite{GoogleScholar2014}\cite{Jacs2006DeflatedIA}, the quality of Google Scholar data is comparable to the data coming from the
subscription-based services for journal publications such as Web of Science \cite{hindex2008}. A Google Scholar page lists the individual's publications with each respective number of citations and overall aggregate citation measures. A profile's aggregate citation measures include the \emph{h}-index and \emph{i10}-index (\emph{i10}). The \emph{h}-index \cite{doi:10.1073/pnas.0507655102} is defined as the highest number \emph{x} for which the individual has \emph{x} number of papers with at least \emph{x} number of citations. The \emph{i10} is the total number of papers with above ten citations by an individual.

We found each page by searching Google Scholar for the professor's name and university. Some professors have common names, and multiple people appeared in the search result. We ensured that the professor's rank, university, and research topics aligned with the page. Sometimes, the Google Scholar page would list a university where the professor was a previous student or faculty member. We confirmed that it was the correct page by looking at past affiliations through their websites or department website pages. 

We found about 89.8\% (5,005/5,574) of professors' Google Scholar pages. About 85.7\% (3,379/3,944) of senior faculty have a page. We determined that using the data of professors who have Google Scholar pages is biased because they tend to have higher citation measures. To prevent bias from affecting the results, we decided to collect citation measures for professors who did not have a page. We introduced the \emph{t10} index, a citation measure that would be easy to collect for professors with or without a Google Scholar page. This measure is explained in the next section. 

\subsection{t10 Index}
The t10 index (\emph{t10}) is defined as the number of citations of one's 10th most cited paper. Identifying this index is more convenient and less prone to error than the \emph{h}-index when performing a manual search. The \emph{t10} is obtained by identifying an individual's ten most cited papers. In contrast, the \emph{h}-index is obtained by identifying the top \emph{x} number of papers that have at least \emph{x} citations. Because assistant professors are starting to build their publication records, we decided to focus on collecting the \emph{t10} of the senior faculty. 

We obtained the \emph{t10} of the associate and full professors with a Google Scholar page using a web scraping program that takes the 10th highest paper on each page. This is a simple process because a Google Scholar page lists the individual's highest-cited articles in descending order. We collected them manually for professors who do not have a Google Scholar page. To save time, we took the \emph{t10} that were gathered in 2017 and matched them to each professor who did not have a \emph{t10}. 

To manually gather the \emph{t10}, we searched Google Scholar for the professor's name. The search engine typically retrieves publications with the name in the author list by descending the number of citations. We looked for the 10th highest cited paper from the search results with the professor's name in the author list. For professors with common names, the search results would show publications from multiple people. We checked each publication to ensure the author was the correct professor. 

We identified the \emph{t10} of 5,553 of the 5,574 CS faculty (99.6\% coverage) and for 3,932 of the 3,944 senior CS faculty (99.7\% coverage) by manually searching Google Scholar. However, when a faculty has a common name or a faculty name listed on the people pages does not precisely match the name listed in their papers, obtaining \emph{t10}-index can be too time-consuming. To save time, the curators were instructed to abort the extraction if it took more than 5 minutes. As a result, we did not collect t10 for 21 of the 5,574 CS faculty (0.4\%). Since a faculty’s name should not have an influence on their citation count, the resulting sample of faculty with known \emph{t10} can be treated as an unbiased sample of the senior CS faculty.

Furthermore, among 3,416 faculty in both the 2017 and 2022 data, we found that 2,520 (73.8\% coverage) have an h-index and \emph{t10}. 459 of them have added google scholar profiles since 2017, and none of them removed their google scholar profile. 926 of them were promoted during the past few years. Among 1,750 new faculty in our 2022 data, 1,636 have a google scholar profile, where 205 are full professors, 196 are associate professors, and 1,235 are assistant professors. 1,741 of these new faculty have \emph{t10}, where 252 are full professors, 219 are associate professors, and 1,270 are assistant professors. 

\section{Methods}
\subsection{Program Strength Measures}
We propose two approaches for using individual citation measures to calculate the strength of a program, averaged and cumulative citation measures, the same as those used in 2017.

\subsection{Averaged citation measures}
The first method we use to measure the strength of a program is by averaging citations of its faculty members \cite{Lazaridis2009RankingUD}. We use three different averaging schemes. The first averaging scheme is the median of \emph{t10} values of senior CS faculty, which we denote as \emph{m10}. The second is the geometric mean of (1+\emph{t10}) values of senior CS faculty, which we denote as \emph{g10}. The third is the average percentile of the senior faculty's \emph{t10}, which we denote as \emph{p10}. We exclude assistant professors from the averaged measures because their citations are typically smaller, and their inclusion would hurt departments with more assistant professors.

\subsection{Cumulative citation measures}
The second method to measure the strength of a program is to count the number of highly cited faculty in a program. We define a t10 threshold to determine which professors are highly cited. We introduce \emph{cN}, which denotes the number of faculty whose t10 is higher than \emph{N}\% of senior faculty, with 0 < \emph{N} < 100. For example, \emph{c40} is the count of professors within a department with a \emph{t10} higher than 40\% of senior faculty. We include all faculty to calculate \emph{cN}.

\subsection{Regression Models}
We use regression models that combine the averaged and cumulative citation measures into an aggregated score. 

The regression models that we consider are of the type in formula \ref{eq:linear regression}:
\begin{equation}
     s_{i} = \beta_{0} + \beta_{1}a_{i} + \beta_{2}c_{i},
     \label{eq:linear regression}
\end{equation}

s\textsubscript{i} is the predicted USN CS score, a\textsubscript{i} is an aggregated citation measure, and c\textsubscript{i} is a cumulative citation measure of the \emph{i}-th program. The regression parameters are \textbeta\textsubscript{0}, \textbeta\textsubscript{1}, and \textbeta\textsubscript{2}. Instead of learning the intercept parameter \textbeta\textsubscript{0}, we set it to \textbeta\textsubscript{0} = 1 by default. The primary justification is that a program with a\textsubscript{i} = 0 and s\textsubscript{i} = 0 does not have active research faculty; based on the peer assessment instructions by USN, this program would have a score of 1 ("marginal"). The secondary justification is that the resulting regression models would be simpler because they only require fitting two regression parameters, \textbeta\textsubscript{1}, and \textbeta\textsubscript{2}. We train one regression model for each combination of the three averaged and the ranges of cumulative measures. We average the individual regression models to create an aggregate score.  

\section{Results}

\subsection{Department size}
\begin{figure}
    \centering
    \includegraphics[width=0.65\textwidth]{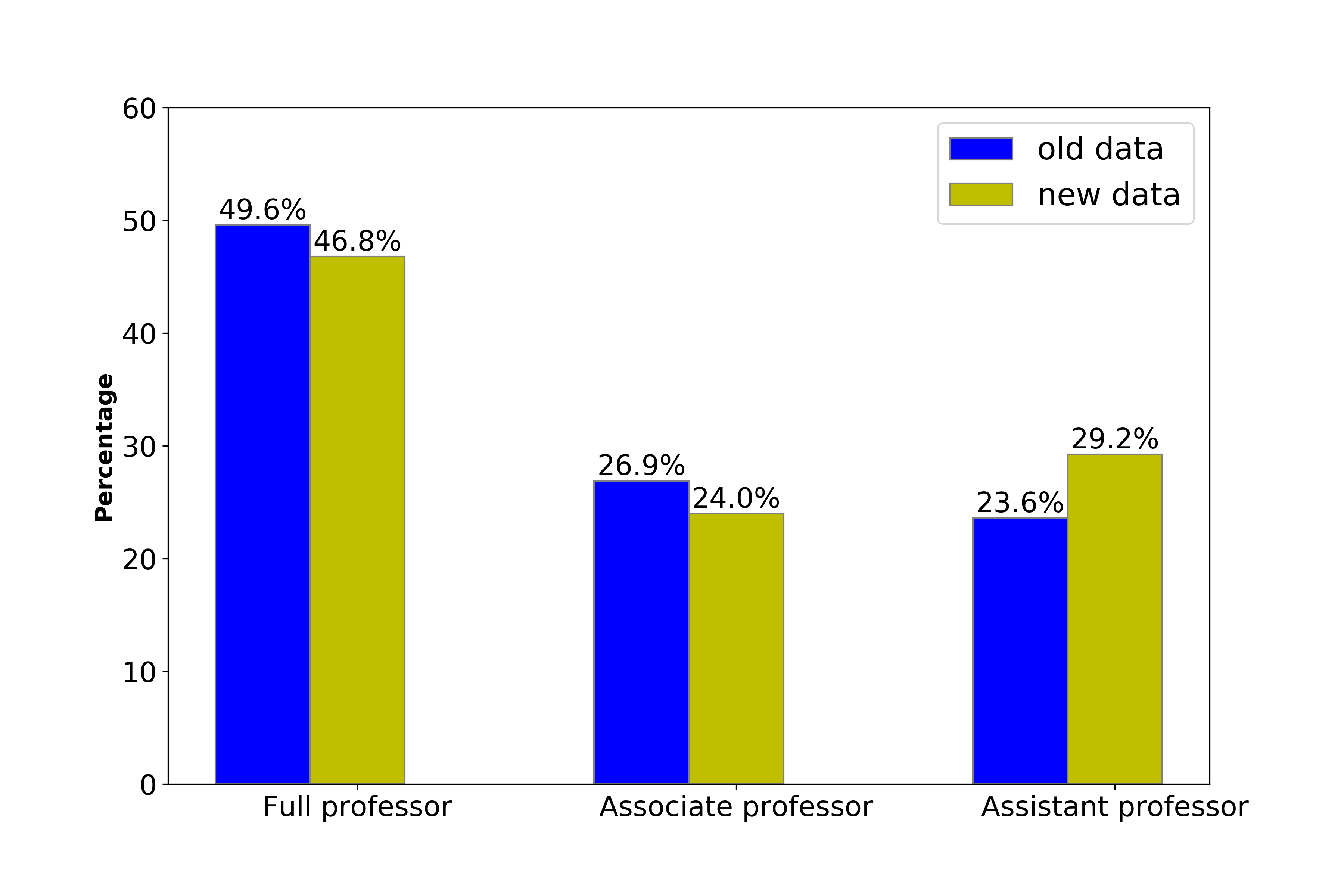}
    \caption{Trend of faculty size}
    \label{fig:size comparison}
\end{figure}

Fig.\ref{fig:size comparison} compares the number of assistant, associate, and full professors between our newly collected data and that from our 2017 paper. The percentage of assistant professors increased during the past four years, which indicates that more young professors are joining CS academia. The reason might be that recently popular areas, such as machine learning or deep learning, are attracting more young professors to contribute to research. 

\begin{figure}
    \centering
    \includegraphics[width=0.65\textwidth]{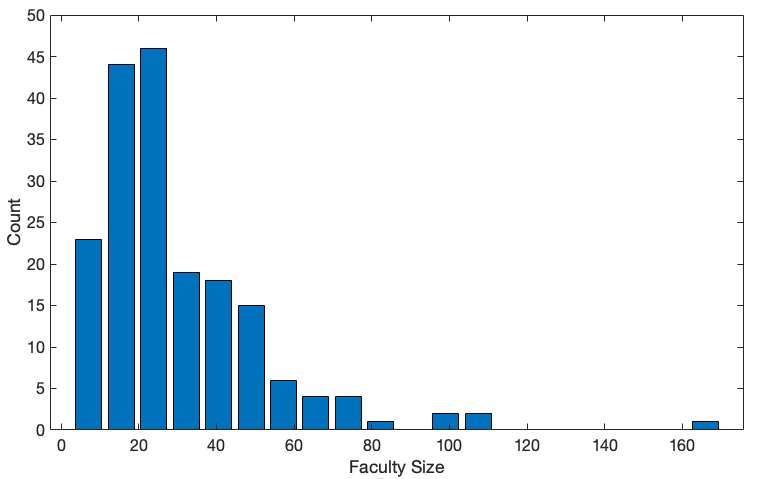}
    \caption{The distribution of the U.S. CS department size}
	\label{fig:department size}
\end{figure}

\begin{figure}
    \centering
    \includegraphics[width=0.65\textwidth]{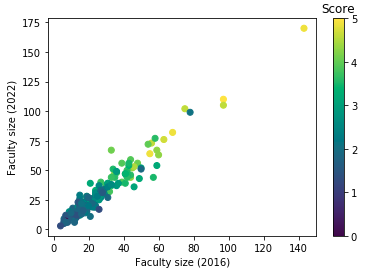}
    \caption{Department size in 2016 and 2022}
	\label{fig:department size compare}
\end{figure}

Fig.\ref{fig:department size} shows the distribution of department sizes, defined as the number of tenure-track faculty in each of the 185 CS programs. The median faculty size is 23, the mode is 20, the minimum is 3, and the maximum is 170 (Carnegie Mellon University). We also show the scatter plot between the department size in 2017 and 2022 in Fig.\ref{fig:department size compare}. The colors represent the USN CS scores, and it can be seen there is an overall increase in department sizes. The department size in 2022 has a similar distribution to that in 2017, since there is a nearly linear relationship between the two. The correlation coefficient of the department sizes and the USN CS scores of the 185 programs is 0.755, which is higher than our previous finding (0.676), indicating that larger departments are more likely to be ranked higher. 

\subsection{Distribution of the Citation Indices}

\begin{figure}
    \centering
    \includegraphics[width=0.65\textwidth]{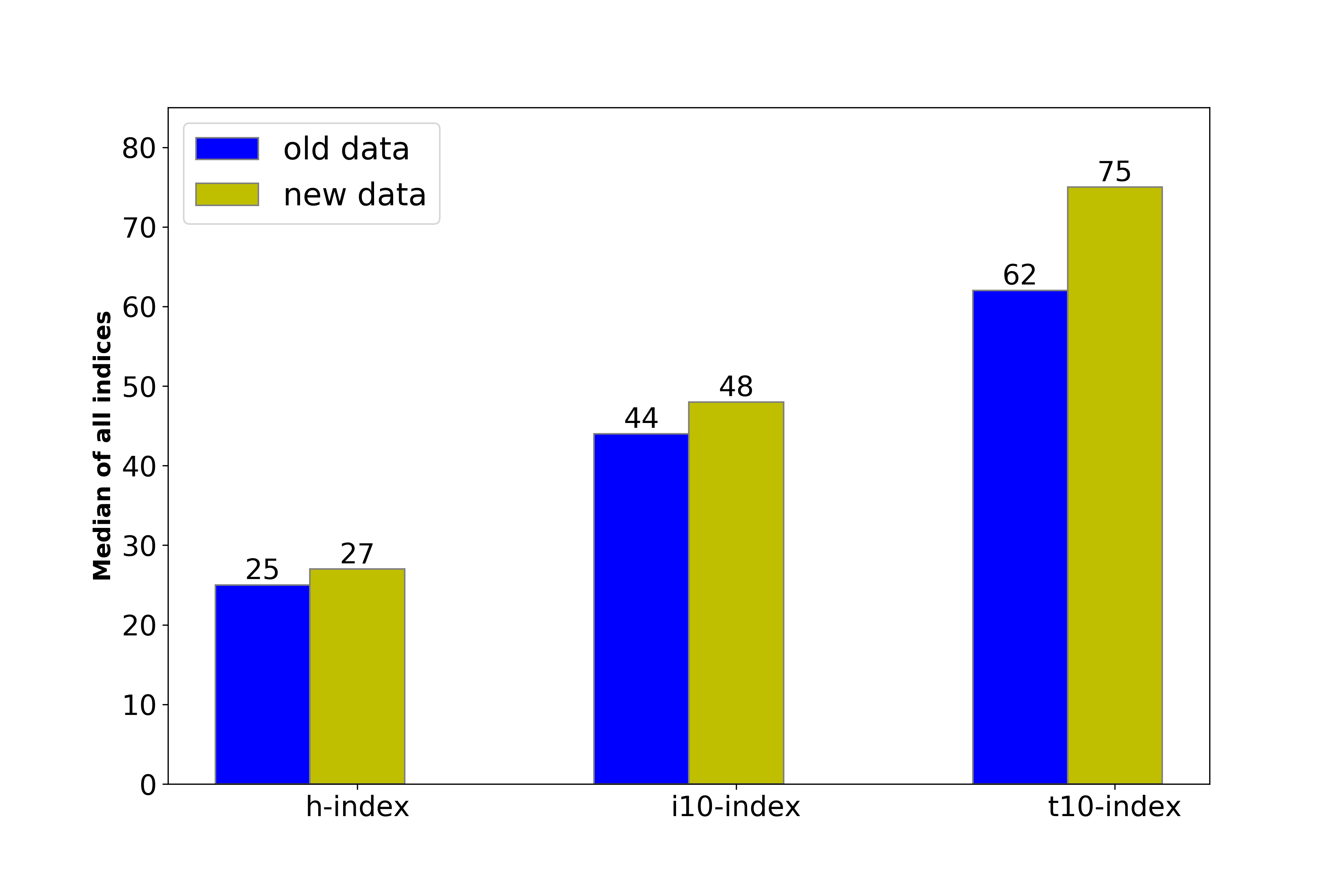}
    \caption{Trend of citation measurements (median value)}
	\label{fig:index comparison}
\end{figure}

As shown in Fig.\ref{fig:index comparison}, the median value of the \emph{h}-index, \emph{i10}-index, and \emph{t10} index all increased compared to our 2017 result. Particularly, the \emph{t10} index has a larger rise compared to the \emph{h}-index and \emph{i10}-index, indicating that it is a more sensitive citation metric. 

\begin{figure}
    \centering
    \includegraphics[width=0.65\textwidth]{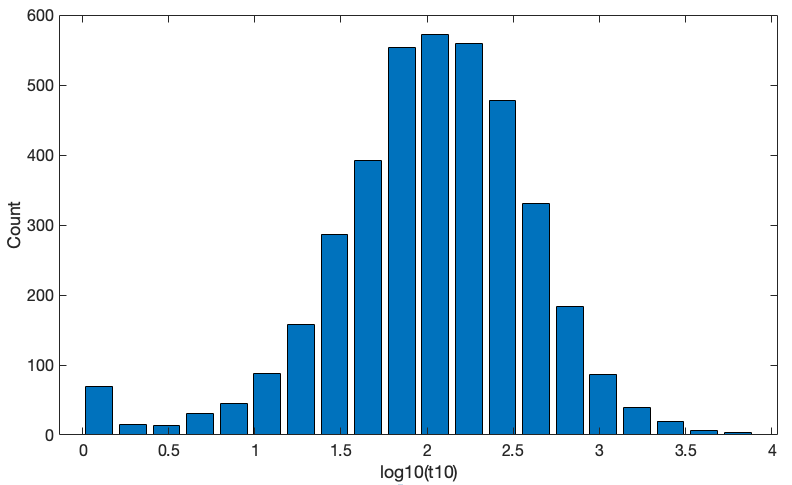}
    \caption{Histogram of t10 of associate and full CS professors}
	\label{fig:t10 hist}
\end{figure}

In Fig.\ref{fig:t10 hist}, we show the histogram of the \emph{t10} for the 3,944 senior faculty. We observe a similar distribution pattern compared to the 2017 result. Since the \emph{t10} distribution resembles a lognormal distribution, the histogram of \emph{t10} is shown in a log scale. We observe a bump at low values, representing the 70 senior faculty with a \emph{t10} of 0, meaning they have less than ten cited papers listed in Google Scholar. The median of \emph{t10} is 114, and the percentiles of \emph{t10} are shown in Table \ref{tab:t10 pecentile}. Overall, the \emph{t10} values increased compared to our results from 2017.

The correlation coefficient between logarithms of \emph{h}-index and \emph{t10} for the 3,379 senior faculty with both indices is 0.943, which is close to our 2017 result. The sufficiently high correlation concludes that the \emph{t10} is a good proxy for the \emph{h}-index.

\begin{table}[h!]
\caption{Percentiles of t10}
\centering
\begin{tabular}{ll}  \hline
           Percentile  & t10   \\ \hline
$10\%$  & 21   \\
$20\%$  & 40   \\
$30\%$  & 60   \\
$40\%$  & 83   \\
$50\%$  & 115   \\
$60\%$  & 154   \\ \hline 

\end{tabular}
\hspace{1em}
\centering
\begin{tabular}{ll}  \hline
           Percentile  & t10   \\ \hline

$70\%$  & 212   \\
$80\%$  & 301   \\
$90\%$  & 493   \\
$95\%$  & 751   \\
$98\%$  & 1247   \\
$99\%$  & 1843   \\    \hline

\end{tabular}
\label{tab:t10 pecentile}
\end{table}

\subsection{Scholar profile bias}
\begin{figure}
    \centering
    \includegraphics[width=0.65\textwidth]{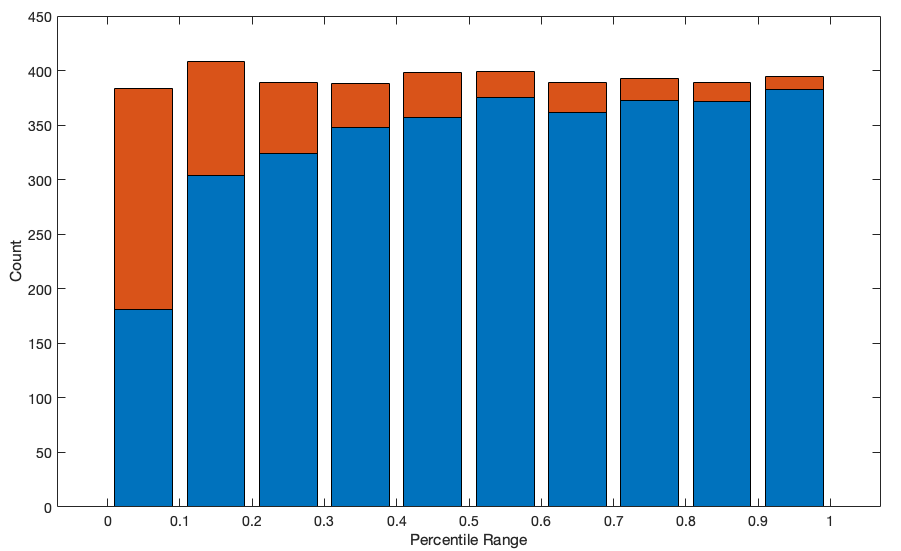}
    \caption{Number of tenured CS faculty with (blue) and without (orange) Google scholar profile as a function of the t10 percentile}
	\label{fig:scholar profile}
\end{figure}

While the median of \emph{t10} for the 3,932 senior CS faculty is 114, it increases to 
131 among the 3,379 professors who have a Google scholar profile and drops to 33 among the 553 without a profile. In Fig. \ref{fig:scholar profile} we show a stacked bar plot of the numbers of faculty with and without Google scholar profiles as a function of their \emph{t10} percentile. These results are consistent with the observations in our 2017 paper, which indicate that CS faculty who have Google scholar profiles are a biased sample of the entire CS faculty and validate our effort to gather the \emph{t10} values and use them in our study instead of the \emph{h}-index. 

\subsection{Scholar model}

According to the above analysis, it can be observed our 2022 data shows a similar pattern and trend compared to our 2017 data. Hence, we decided to keep the same linear regression scholar model used in 2017 and directly applied it to our 2022 data, which is shown in formula \ref{eq:2}:

\begin{equation}
     s = 1 + 0.058\sqrt{m10} + 0.059\sqrt{g10} + 0.121\sqrt{c40} + 0.127\sqrt{c60}
     \label{eq:2}
\end{equation}

\begin{figure}
    \centering
    \includegraphics[width=0.75\textwidth]{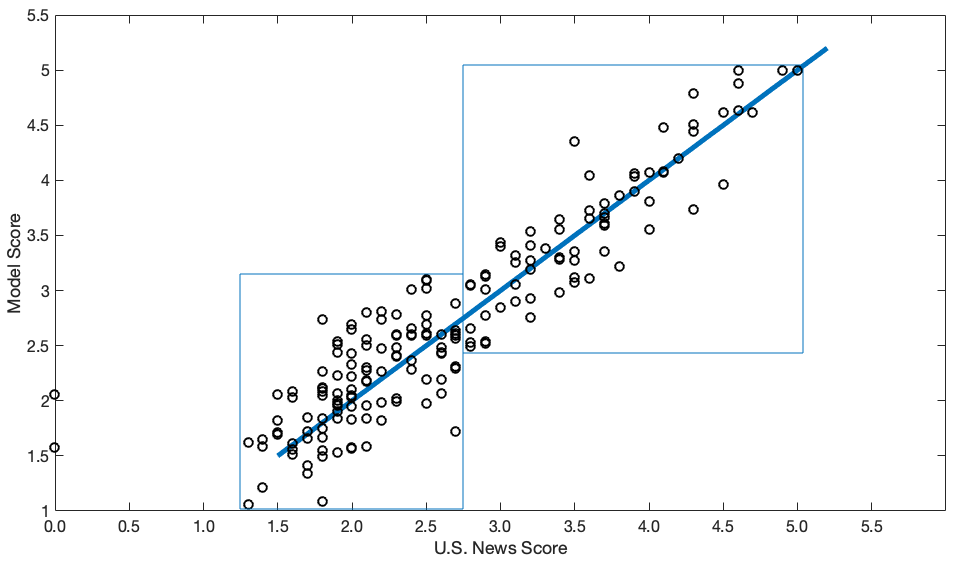}
    \caption{Comparison of scholar model scores and USN CS scores of CS graduate programs. }
	\label{fig:scholar model}
\end{figure}

In Fig \ref{fig:scholar model}, we show a scatter plot of the USN CS scores and scholar model scores for the 185 CS programs. A closer look at the scatter plot reveals that two groups of CS programs can be distinguished with respect to the correlation between the USN CS scores and scholar model scores. The first group contains 73 programs that were scored 2.8 and higher by the USN. The correlation between the USN CS scores and joint model scores in this group is 0.914. The second group contains 112 programs with USN scores between 0 and 2.7. The correlation between the USN CS scores and scholar model scores in this low-scoring group is 0.673, which is much higher than what we observed in 2017 (0.360) but still lower than those 73 programs. We hypothesize that the CS programs whose USN CS scores are between 0 and 2.8 are not sufficiently well-known among the peers to provide objective and reliable peer assessment at the national level.

\subsection{Comparison Study}

We first study the newly added professor list to compare our new ranking result with the previous one conducted in 2017. Among those 1,750 new professors, it is observed that 1,518 ($86.7\%$) of them have a t10-index lower than the average t10-index of all faculty from the department they join in. 223 of them are higher, and 9 of them have a missing t10-index. Furthermore, among those 261 professors who transferred to another department, 114 have joined a higher-ranking department than their previous department, and 147 have joined a lower-ranking department. We also compared their t10-index in 2017 with the previous department average before they moved. It is an interesting finding that only 66 have a t10-index higher than the department average, and 176 have a t10-index lower or equal to the department average. 19 of them have a missing t10-index. Based on these observations, it can be inferred that new professors are primarily young, with a lower t10-index. It is also true that young professors at the starting stage of their careers are more likely to transfer to another university, and most of them tend to join a higher-ranking university. The story might be they built a stronger profile during the past 5 years and then joined a higher-ranking department.

\subsubsection{2017 ranking vs. 2022 ranking}

\begin{figure}
    \centering
    \includegraphics[width=0.55\textwidth]{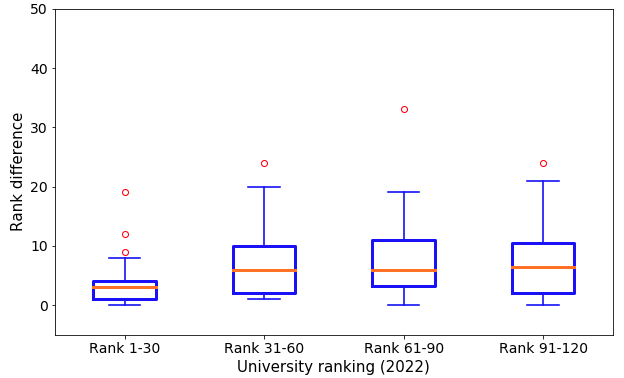}
    \caption{Absolute value of ranking difference between our 2017 and 2022 scholar ranking results}
	\label{fig:scholar diff}
\end{figure}

To better understand how the ranking has changed during the past five years, we calculate the ranking difference between the results we produced in 2017 and 2022 using the same scholar model created in 2017. Fig. \ref{fig:scholar diff} shows the box plot of the absolute difference between our old (2017) and new (2022) scholar ranking results. It can be seen that the placements of top universities above rank 90 are more stable, while lower-ranked universities tend to have more significant variations. However, we observe some outliers. For example, Northeastern University jumped from rank 40 to 21 because this department may have recruited many new professors since 2017. 

Another observation is that larger departments are not necessarily ranked higher. For example, Stanford University and Princeton University both have relatively small departments. Still, they are within the top 10 departments in our ranking, indicating that other factors, such as faculty citations, significantly influence the ranking result. 

\begin{figure}
    \centering
    \includegraphics[width=0.55\textwidth]{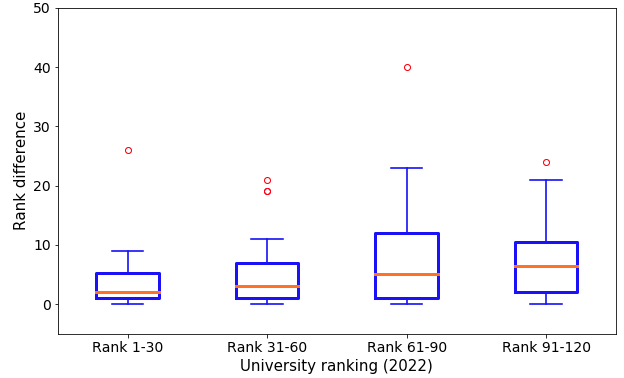}
    \caption{Absolute value of ranking difference between USN 2017 and USN 2022 results}
	\label{fig:usn diff}
\end{figure}

Fig. \ref{fig:usn diff} shows the absolute difference between the USN 2017 and USN 2022 ranking. It can be observed that the ranking difference has a similar pattern compared to what has been shown in Fig. \ref{fig:scholar diff}. This indicates that our method using the t-10 index of faculty to rank the program produces similar ranking results compared to USN. 

\subsubsection{Scholar ranking vs. USN}

The new ranking results and a comparison to the USN scores are shown in the Appendix. To justify that it is appropriate to apply the linear regression model obtained from the 2017 data to the 2022 data, we calculated the correlations between our scholar ranking result and the USN ranking in 2017 and 2022, respectively. The results are shown in Table \ref{tab:usn_vs_scholar}.

\begin{table}[h!]
\centering
\caption{Correlation between USN ranking and our scholar ranking using the 2017 regression model}
\hspace{1em}
\setlength{\tabcolsep}{20pt}
 \begin{tabular}{|c|c|c|c|} 
 \hline
                & $R^2$    & Pearson & Spearman \\ \hline
USN 2017 vs. scholar ranking 2017        & 0.8731   & 0.9357   & 0.8978  \\ \hline
USN 2022 vs. scholar ranking 2022           & 0.8734   & 0.9390 & 0.9126 \\ \hline

\end{tabular}
 \label{tab:usn_vs_scholar}
\end{table}

Despite the fact that we used the same regression formula for both 2017 and 2022 data, our scholar ranking results show a high correlation with both USN 2017 and USN 2022 rankings. This result confirms that our practice of using the 2017 ranking formula on our 2022 data is justifiable.

\begin{figure}
    \centering
    \includegraphics[width=0.55\textwidth]{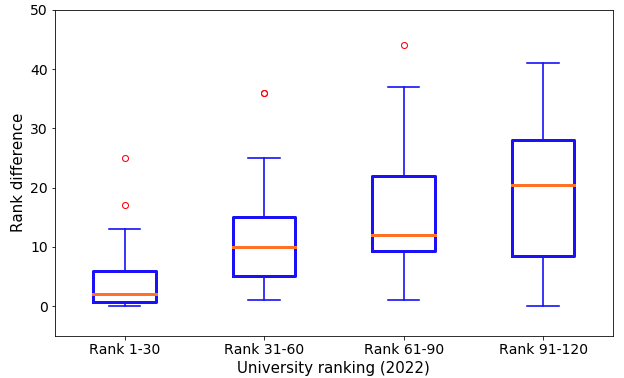}
    \caption{Absolute value of ranking difference between our 2022 scholar ranking and USN 2022 ranking}
	\label{fig:usn_scholar_diff}
\end{figure}

To further investigate the relationship between our scholar ranking and the USN ranking, we calculated the difference between our new ranking and the USN 2022 ranking. Fig. \ref{fig:usn_scholar_diff} shows the boxplot of the absolute value of difference. We separate universities into six groups based on their new scholar rank. It can be observed that our ranking model is more likely to match with the USN ranking for higher-ranking departments. For instance, all departments that rank 1-30 show a rank difference of less than 10. 

\begin{figure}
    \centering
    \includegraphics[width=0.55\textwidth]{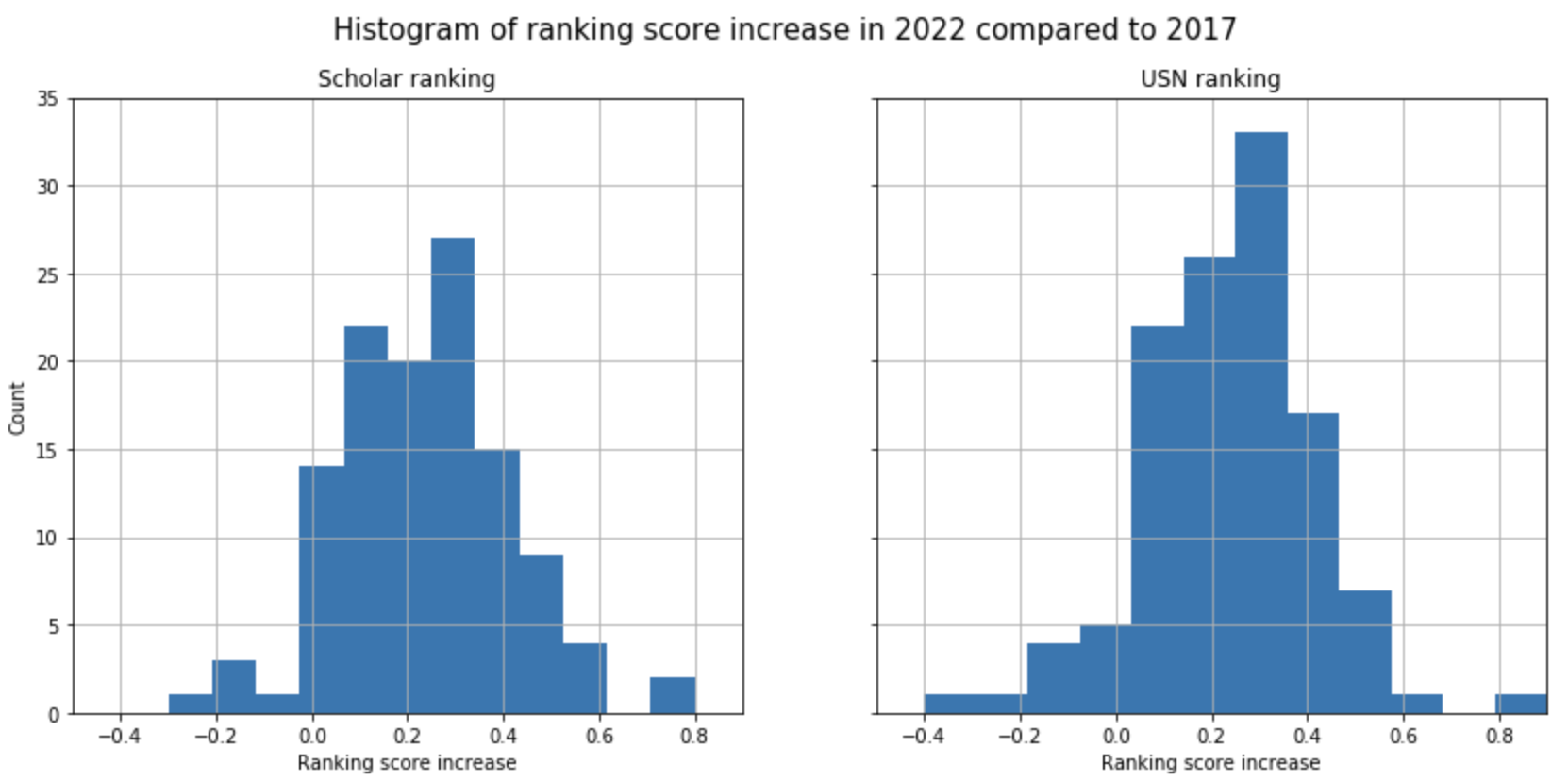}
    \caption{Histogram plot of ranking score increase between 2022 and 2017}
	\label{fig:score_increase}
\end{figure}

We also calculate the ranking score increase in 2022 compared to 2017 for both our scholar model and USN and show the histogram plots in Fig. \ref{fig:score_increase}. It can be seen that most departments have a score increase in both the scholar model and USN, and all of them are within the range between -0.4 and 0.8. There are a few extreme cases. For example, the ranking score of Northeastern University increased by 0.8; where the reason may be the department recruited many new faculty during the past few years. UNC Chapel Hill has a score decrease of 0.3. Their department size remains the same (32), but the \emph{m10}, \emph{g10}, \emph{c40}, and \emph{c60} values drop in 2022.

\subsubsection{Scholar ranking vs. CSRankings}

\begin{table}[h!]
\centering
\caption{Correlation between CSRankings, USN 2022 and our scholar ranking}
\setlength{\tabcolsep}{20pt}
 \begin{tabular}{|c|c|c|c|} 
 \hline
                & $R^2$    & Pearson & Spearman \\ \hline
USN vs. scholar ranking        & \bf{0.8734}   & \bf{0.9390}   & 0.9126  \\ \hline
USN vs. CSRankings            & 0.8391   & 0.9160 & 0.9157 \\ \hline
scholar ranking vs. CSRankings  & 0.8375   & 0.9151 & 0.8965 \\ \hline
USN vs. average model      & 0.8462   & 0.9199 & \bf{0.9305} \\ \hline
\end{tabular}
 \label{tab:csrankings_vs_scholar}
\end{table}

\begin{figure}
    \centering
    \includegraphics[width=0.55\textwidth]{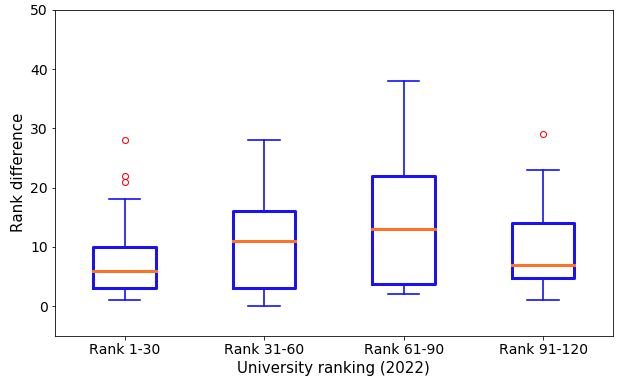}
    \caption{Absolute value of ranking difference between CSRankings and USN 2022}
	\label{fig:csrankings usn}
\end{figure}

\begin{figure}
    \centering
    \includegraphics[width=0.55\textwidth]{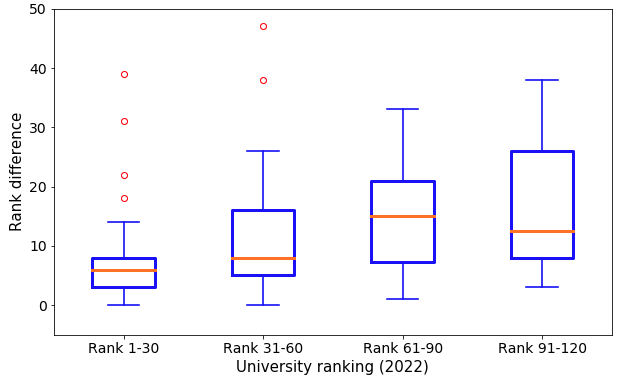}
    \caption{Absolute value of ranking difference between CSRankings and our scholar ranking}
	\label{fig:csrankings scholar}
\end{figure}

To further investigate the relationship between our ranking and some widely-used CS ranking results, we compared the CSRankings result with our ranking. Unlike USN ranking, CSRankings\footnote{\url{http://csrankings.org}} relies on publications in top-tier computer science conferences, as reported by DBLP, a computer science bibliography website \footnote{\url{https://dblp.org/}}. To study the relationship between CSRankings and our scholar ranking, we collected the current CSRankings result and calculated its correlations with USN and our scholar ranking. Since the CSRanking score is based on a different scale, we applied log transformation to its score before computing the correlations. The result is shown in Table \ref{tab:csrankings_vs_scholar}. 

It can be seen from the Table that our scholar ranking result has a higher correlation with the USN ranking, which indicates that it is better aligned with the USN ranking compared to CSRankings. This indication is further justified in Fig. \ref{fig:usn_scholar_diff} and Fig. \ref{fig:csrankings usn}, where we show the ranking difference between 1) our scholar ranking results with USN and 2) CSRankings with USN. It can be seen that CSRankings shows an overall more significant ranking difference with USN compared to our scholar model ranking, especially for the top 60 departments. An interesting finding is that CSRankings is better aligned with USN than our scholar ranking for those lower-ranking departments (>90). The ranking difference between CSRankings and our scholar model ranking is shown in Fig. \ref{fig:csrankings scholar}. The result shows that the difference increases with a larger variance as the ranking of the department decreases. This indicates that the ranking of top departments is more stable despite which model is being used, whereas lower-ranking departments are more sensitive to the ranking method. We also build an average model by computing the average score using the scholar model score and CSRankings score (shown in the last row of Table), which yields a higher spearman correlation with the USN ranking. 

\bibliographystyle{unsrt}
\bibliography{template}   


\clearpage
\appendix


{
\setlength{\extrarowheight}{0pt}
\addtolength{\extrarowheight}{\aboverulesep}
\addtolength{\extrarowheight}{\belowrulesep}
\setlength{\aboverulesep}{0pt}
\setlength{\belowrulesep}{0pt}
\fontsize{8}{10}\selectfont
\begin{longtable}{lllllllllll}
\caption{List of 185 U.S. CS graduate programs: Ranking by our scholar model (Rank), University name (University), Number of tenured faculty with t10 score (Size), median t10 score of all the faculty (M10), geometric mean of t10 score of all faculty (G10), number of highly cited faculty based on c40 (C40) and c60 (C60), U.S. News CS score (USN), Scholar score (Scholar)} \\
\toprule
\textbf{Rank}                                  & \textbf{University}                           & \textbf{Size} & \textbf{M10} & \textbf{G10} & \textbf{C40} & \textbf{C60} & \textbf{USN} & \textbf{Scholar} & \textbf{USN 2016} & \textbf{Scholar 2016}  \endfirsthead
\rowcolor[rgb]{0.733,0.827,0.906} \textbf{1}   & Carnegie Mellon University                    & 170           & 280          & 262          & 121          & 84           & 4.9          & 5                & 5                 & 5                      \\
\textbf{1}                                     & Cornell University                            & 102           & 327          & 315          & 75           & 58           & 4.6          & 5                & 4.5               & 4.4                    \\
\rowcolor[rgb]{0.733,0.827,0.906} \textbf{1}   & Massachusetts Institute of Technology         & 110           & 304          & 302          & 92           & 74           & 5            & 5                & 5                 & 5                      \\
\textbf{1}                                     & Stanford University                           & 64            & 707          & 706          & 60           & 53           & 4.9          & 5                & 5                 & 5                      \\
\rowcolor[rgb]{0.733,0.827,0.906} \textbf{1}   & University of California Berkeley             & 82            & 421          & 461          & 68           & 60           & 4.9          & 5                & 5                 & 5                      \\
\textbf{6}                                     & University of Washington                      & 73            & 345          & 295          & 58           & 47           & 4.6          & 4.9              & 4.5               & 4.3                    \\
\rowcolor[rgb]{0.733,0.827,0.906} \textbf{7}   & University of California San Diego            & 63            & 341          & 330          & 49           & 39           & 4.3          & 4.8              & 4                 & 4.2                    \\
\textbf{8}                                     & Georgia Institute of Technology               & 105           & 203          & 198          & 75           & 53           & 4.6          & 4.6              & 4.3               & 4.3                    \\
\rowcolor[rgb]{0.733,0.827,0.906} \textbf{8}   & Princeton University                          & 48            & 360          & 345          & 36           & 30           & 4.5          & 4.6              & 4.4               & 4.1                    \\
\textbf{8}                                     & University of Illinois Urbana Champaign       & 76            & 265          & 248          & 56           & 43           & 4.7          & 4.6              & 4.6               & 4.1                    \\
\rowcolor[rgb]{0.733,0.827,0.906} \textbf{11}  & University of California Los Angeles          & 44            & 317          & 316          & 35           & 31           & 4.3          & 4.5              & 4.1               & 4.2                    \\
\textbf{11}                                    & University of Pennsylvania                    & 67            & 256          & 238          & 49           & 39           & 4.1          & 4.5              & 3.8               & 3.7                    \\
\rowcolor[rgb]{0.733,0.827,0.906} \textbf{13}  & Columbia University                           & 52            & 249          & 288          & 42           & 34           & 4.3          & 4.4              & 4                 & 4.1                    \\
\textbf{13}                                    & New York University                           & 44            & 288          & 254          & 38           & 29           & 3.5          & 4.4              & 3.4               & 4                      \\
\rowcolor[rgb]{0.733,0.827,0.906} \textbf{13}  & University of Michigan Ann Arbor              & 67            & 251          & 254          & 47           & 35           & 4.3          & 4.4              & 4.1               & 4.1                    \\
\textbf{16}                                    & Harvard University                            & 34            & 277          & 286          & 28           & 22           & 4.2          & 4.2              & 3.9               & 3.7                    \\
\rowcolor[rgb]{0.733,0.827,0.906} \textbf{17}  & Duke University                               & 56            & 216          & 195          & 37           & 26           & 3.9          & 4.1              & 3.6               & 3.6                    \\
\textbf{17}                                    & Johns Hopkins University                      & 32            & 258          & 309          & 23           & 17           & 4            & 4.1              & 3.5               & 4                      \\
\rowcolor[rgb]{0.733,0.827,0.906} \textbf{17}  & University of Maryland College Park           & 56            & 219          & 186          & 37           & 29           & 4.1          & 4.1              & 4                 & 4                      \\
\textbf{17}                                    & University of Wisconsin Madison               & 45            & 285          & 225          & 28           & 20           & 4.1          & 4.1              & 4.2               & 3.9                    \\
\rowcolor[rgb]{0.733,0.827,0.906} \textbf{21}  & Northeastern University                       & 77            & 165          & 180          & 45           & 30           & 3.6          & 4                & 2.7               & 3.2                    \\
\textbf{21}                                    & University of Southern California             & 40            & 272          & 238          & 26           & 19           & 3.9          & 4                & 3.7               & 3.9                    \\
\rowcolor[rgb]{0.733,0.827,0.906} \textbf{21}  & University of Texas Austin                    & 53            & 182          & 184          & 38           & 25           & 4.5          & 4                & 4.3               & 3.7                    \\
\textbf{24}                                    & Brown University                              & 31            & 224          & 233          & 22           & 17           & 3.8          & 3.9              & 3.7               & 3.5                    \\
\rowcolor[rgb]{0.733,0.827,0.906} \textbf{24}  & University of Massachusetts Amherst           & 59            & 191          & 187          & 33           & 22           & 3.9          & 3.9              & 3.6               & 3.7                    \\
\textbf{26}                                    & University of Chicago                         & 48            & 174          & 208          & 29           & 17           & 3.7          & 3.8              & 3.3               & 3.5                    \\
\rowcolor[rgb]{0.733,0.827,0.906} \textbf{26}  & Yale University                               & 26            & 243          & 231          & 19           & 14           & 4            & 3.8              & 3.7               & 4                      \\
\textbf{28}                                    & California Institute of Technology            & 21            & 240          & 240          & 16           & 12           & 4.3          & 3.7              & 4.2               & 3.7                    \\
\rowcolor[rgb]{0.733,0.827,0.906} \textbf{28}  & Pennsylvania State University University Park & 44            & 200          & 176          & 25           & 17           & 3.6          & 3.7              & 3.4               & 3.4                    \\
\textbf{28}                                    & Rice University                               & 28            & 213          & 204          & 18           & 13           & 3.7          & 3.7              & 3.7               & 3.3                    \\
\rowcolor[rgb]{0.733,0.827,0.906} \textbf{28}  & University of California Santa Barbara        & 37            & 183          & 194          & 25           & 15           & 3.7          & 3.7              & 3.3               & 3.6                    \\
\textbf{28}                                    & University of Minnesota Twin Cities           & 46            & 145          & 186          & 30           & 15           & 3.6          & 3.7              & 3.4               & 3.4                    \\
\rowcolor[rgb]{0.733,0.827,0.906} \textbf{33}  & Purdue University West Lafayette              & 72            & 137          & 133          & 31           & 17           & 4            & 3.6              & 3.7               & 3.3                    \\
\textbf{33}                                    & Stony Brook University SUNY                   & 45            & 152          & 137          & 27           & 17           & 3.4          & 3.6              & 3.1               & 3.3                    \\
\rowcolor[rgb]{0.733,0.827,0.906} \textbf{33}  & University of California Davis                & 37            & 178          & 162          & 23           & 18           & 3.4          & 3.6              & 3.3               & 3.5                    \\
\textbf{33}                                    & University of California Irvine               & 50            & 152          & 132          & 28           & 19           & 3.7          & 3.6              & 3.4               & 3.4                    \\
\rowcolor[rgb]{0.733,0.827,0.906} \textbf{33}  & University of Virginia                        & 40            & 193          & 179          & 20           & 14           & 3.7          & 3.6              & 3.4               & 3.1                    \\
\textbf{38}                                    & University of California Santa Cruz           & 39            & 185          & 153          & 20           & 14           & 3.2          & 3.5              & 2.8               & 3.5                    \\
\rowcolor[rgb]{0.733,0.827,0.906} \textbf{39}  & Boston University                             & 33            & 141          & 167          & 18           & 11           & 3.3          & 3.4              & 3                 & 3.2                    \\ 
\textbf{39}                                    & Michigan State University                     & 39            & 151          & 159          & 20           & 12           & 3            & 3.4              & 2.8               & 3                      \\ 
\rowcolor[rgb]{0.733,0.827,0.906} \textbf{39}  & Northwestern University                       & 44            & 122          & 123          & 26           & 12           & 3.7          & 3.4              & 3.3               & 3.1                    \\
\textbf{39}                                    & Rutgers University                            & 39            & 151          & 152          & 18           & 10           & 3.5          & 3.4              & 3.3               & 3.3                    \\
\rowcolor[rgb]{0.733,0.827,0.906} \textbf{39}  & University of Arizona                         & 22            & 197          & 168          & 14           & 9            & 3.2          & 3.4              & 3.1               & 3.2                    \\
\textbf{39}                                    & University of California Riverside            & 34            & 153          & 133          & 22           & 12           & 3            & 3.4              & 2.8               & 3.3                    \\
\rowcolor[rgb]{0.733,0.827,0.906} \textbf{45}  & Arizona State University                      & 54            & 92           & 122          & 21           & 16           & 3.2          & 3.3              & 3                 & 2.9                    \\
\textbf{45}                                    & University of Colorado Boulder                & 51            & 127          & 100          & 24           & 12           & 3.5          & 3.3              & 3.1               & 3                      \\
\rowcolor[rgb]{0.733,0.827,0.906} \textbf{45}  & University of Rochester                       & 18            & 186          & 160          & 10           & 7            & 3.1          & 3.3              & 2.9               & 3                      \\
\textbf{45}                                    & University of Utah                            & 54            & 117          & 121          & 22           & 12           & 3.4          & 3.3              & 3.1               & 3                      \\
\rowcolor[rgb]{0.733,0.827,0.906} \textbf{45}  & Vanderbilt University                         & 26            & 164          & 155          & 13           & 10           & 3.1          & 3.3              & 2.8               & 2.9                    \\
\textbf{45}                                    & Washington University in St Louis             & 28            & 135          & 150          & 17           & 10           & 3.4          & 3.3              & 3.1               & 2.9                    \\
\rowcolor[rgb]{0.733,0.827,0.906} \textbf{51}  & University of North Carolina Chapel Hill      & 32            & 139          & 122          & 16           & 10           & 3.8          & 3.2              & 3.6               & 3.5                    \\
\textbf{51}                                    & University of Notre Dame                      & 27            & 126          & 144          & 17           & 7            & 3.2          & 3.2              & 2.7               & 2.7                    \\
\rowcolor[rgb]{0.733,0.827,0.906} \textbf{53}  & Colorado State University                     & 22            & 115          & 136          & 14           & 7            & 2.5          & 3.1              & 2.4               & 3                      \\
\textbf{53}                                    & George Mason University                       & 47            & 100          & 99           & 21           & 10           & 2.9          & 3.1              & 2.5               & 2.9                    \\
\rowcolor[rgb]{0.733,0.827,0.906} \textbf{53}  & Indiana University Bloomington                & 36            & 103          & 99           & 17           & 9            & 3.1          & 3.1              & 2.9               & 3                      \\
\textbf{53}                                    & Ohio State University                         & 44            & 99           & 84           & 22           & 11           & 3.6          & 3.1              & 3.3               & 3.1                    \\
\rowcolor[rgb]{0.733,0.827,0.906} \textbf{53}  & Texas AM University College Station           & 48            & 94           & 90           & 22           & 9            & 3.5          & 3.1              & 3.1               & 2.9                    \\
\textbf{53}                                    & University of Central Florida                 & 37            & 89           & 100          & 18           & 10           & 2.8          & 3.1              & 2.2               & 2.6                    \\
\rowcolor[rgb]{0.733,0.827,0.906} \textbf{53}  & University of Tennessee Knoxville             & 29            & 125          & 114          & 13           & 9            & 2.5          & 3.1              & 2.4               & 3                      \\
\textbf{53}                                    & University of Texas Dallas                    & 52            & 88           & 76           & 24           & 15           & 2.9          & 3.1              & 2.4               & 2.9                    \\
\rowcolor[rgb]{0.733,0.827,0.906} \textbf{53}  & Virginia Tech                                 & 56            & 93           & 99           & 24           & 9            & 3.5          & 3.1              & 3.1               & 3                      \\
\textbf{62}                                    & College of William and Mary                   & 21            & 136          & 140          & 9            & 6            & 2.8          & 3                & 2.4               & 2.8                    \\
\rowcolor[rgb]{0.733,0.827,0.906} \textbf{62}  & Lehigh University                             & 20            & 148          & 129          & 9            & 5            & 2.5          & 3                & 2.1               & 2.7                    \\
\textbf{62}                                    & Temple University                             & 23            & 130          & 103          & 12           & 7            & 2.4          & 3                & 2                 & 2.6                    \\
\rowcolor[rgb]{0.733,0.827,0.906} \textbf{62}  & University at Buffalo SUNY                    & 39            & 96           & 90           & 17           & 9            & 2.9          & 3                & 2.6               & 3                      \\
\textbf{62}                                    & University of Florida                         & 47            & 82           & 79           & 18           & 11           & 3.4          & 3                & 3                 & 2.7                    \\
\rowcolor[rgb]{0.733,0.827,0.906} \textbf{67}  & North Carolina State University               & 43            & 82           & 83           & 16           & 9            & 3.2          & 2.9              & 3                 & 2.9                    \\
\textbf{67}                                    & Rensselaer Polytechnic Institute              & 19            & 89           & 134          & 9            & 6            & 3.1          & 2.9              & 2.9               & 2.8                    \\
\rowcolor[rgb]{0.733,0.827,0.906} \textbf{67}  & University of Maryland Baltimore County       & 29            & 94           & 101          & 12           & 6            & 2.7          & 2.9              & 2.4               & 2.7                    \\
\textbf{67}                                    & University of Pittsburgh                      & 23            & 104          & 104          & 11           & 6            & 3.1          & 2.9              & 2.9               & 2.8                    \\
\rowcolor[rgb]{0.733,0.827,0.906} \textbf{71}  & CUNY Graduate School and University Center    & 99            & 37           & 43           & 25           & 13           & 2.1          & 2.8              & 2.3               & 2.6                    \\
\textbf{71}                                    & Dartmouth College                             & 20            & 97           & 110          & 8            & 3            & 3.2          & 2.8              & 3.1               & 2.7                    \\
\rowcolor[rgb]{0.733,0.827,0.906} \textbf{71}  & Oregon State University                       & 49            & 84           & 69           & 17           & 4            & 2.9          & 2.8              & 2.5               & 2.3                    \\
\textbf{71}                                    & University of Houston                         & 24            & 95           & 88           & 13           & 4            & 2.2          & 2.8              & 2.1               & 2.4                    \\
\rowcolor[rgb]{0.733,0.827,0.906} \textbf{71}  & University of Illinois Chicago                & 39            & 85           & 100          & 13           & 5            & 3            & 2.8              & 2.7               & 2.7                    \\
\textbf{71}                                    & University of Texas Arlington                 & 37            & 85           & 69           & 13           & 6            & 2.5          & 2.8              & 2.2               & 2.7                    \\
\rowcolor[rgb]{0.733,0.827,0.906} \textbf{71}  & Wayne State University                        & 22            & 109          & 89           & 11           & 3            & 2.3          & 2.8              & 2                 & 2.4                    \\
\textbf{78}                                    & New Jersey Institute of Technology            & 30            & 70           & 70           & 11           & 6            & 2.5          & 2.7              & 2.2               & 2.4                    \\
\rowcolor[rgb]{0.733,0.827,0.906} \textbf{78}  & Portland State University                     & 19            & 94           & 71           & 7            & 6            & 2            & 2.7              & 0                 & 2.7                    \\
\textbf{78}                                    & Tufts University                              & 21            & 81           & 95           & 8            & 3            & 2.8          & 2.7              & 2.4               & 2.4                    \\
\rowcolor[rgb]{0.733,0.827,0.906} \textbf{78}  & University of Memphis                         & 18            & 88           & 99           & 7            & 5            & 1.8          & 2.7              & 0                 & 2.4                    \\
\textbf{78}                                    & University of New Mexico                      & 16            & 93           & 89           & 7            & 3            & 2.4          & 2.7              & 2.3               & 2.4                    \\
\rowcolor[rgb]{0.733,0.827,0.906} \textbf{78}  & University of California Merced               & 20            & 91           & 106          & 7            & 4            & 2.2          & 2.7              &                   &                        \\
\textbf{84}                                    & Binghamton University SUNY                    & 27            & 76           & 56           & 11           & 4            & 2.4          & 2.6              & 2                 & 2.2                    \\
\rowcolor[rgb]{0.733,0.827,0.906} \textbf{84}  & Drexel University                             & 15            & 94           & 80           & 6            & 2            & 2.7          & 2.6              & 2.2               & 2.4                    \\
\textbf{84}                                    & Illinois Institute of Technology              & 23            & 74           & 65           & 9            & 4            & 2.4          & 2.6              & 2.1               & 2.5                    \\
\rowcolor[rgb]{0.733,0.827,0.906} \textbf{84}  & University of Connecticut                     & 28            & 102          & 78           & 9            & 1            & 2.7          & 2.6              & 2.3               & 2.3                    \\
\textbf{84}                                    & University of Georgia                         & 24            & 81           & 65           & 8            & 4            & 2.5          & 2.6              & 2.2               & 2.2                    \\
\rowcolor[rgb]{0.733,0.827,0.906} \textbf{84}  & University of Massachusetts Lowell            & 26            & 90           & 47           & 7            & 5            & 2.1          & 2.6              & 0                 & 2.1                    \\
\textbf{84}                                    & University of Missouri                        & 29            & 62           & 65           & 11           & 4            & 2.3          & 2.6              & 2.1               & 2.4                    \\
\rowcolor[rgb]{0.733,0.827,0.906} \textbf{84}  & University of Nebraska Lincoln                & 31            & 78           & 68           & 10           & 3            & 2.6          & 2.6              & 2.4               & 2.6                    \\
\textbf{84}                                    & University of Oregon                          & 20            & 97           & 93           & 7            & 2            & 2.7          & 2.6              & 2.6               & 2.2                    \\
\rowcolor[rgb]{0.733,0.827,0.906} \textbf{84}  & University of South Florida                   & 26            & 60           & 77           & 7            & 6            & 2.3          & 2.6              & 2.1               & 2.5                    \\
\textbf{84}                                    & Washington State University                   & 22            & 75           & 85           & 8            & 3            & 2.7          & 2.6              & 2.4               & 2                      \\
\rowcolor[rgb]{0.733,0.827,0.906} \textbf{84}  & West Virginia University                      & 11            & 117          & 62           & 6            & 4            & 2            & 2.6              & 2                 & 2.3                    \\
\textbf{84}                                    & Worcester Polytechnic Institute               & 32            & 69           & 76           & 9            & 4            & 2.5          & 2.6              & 2.2               & 2.4                    \\
\rowcolor[rgb]{0.733,0.827,0.906} \textbf{97}  & Case Western Reserve University               & 16            & 91           & 68           & 7            & 2            & 2.9          & 2.5              & 2.4               & 2.4                    \\
\textbf{97}                                    & Florida Atlantic University                   & 27            & 53           & 55           & 11           & 4            & 1.9          & 2.5              & 0                 & 2.1                    \\
\rowcolor[rgb]{0.733,0.827,0.906} \textbf{97}  & Florida International University              & 34            & 60           & 42           & 12           & 4            & 2.1          & 2.5              & 0                 & 2.2                    \\
\textbf{97}                                    & Georgia State University                      & 28            & 59           & 65           & 9            & 3            & 2.1          & 2.5              & 2                 & 2.3                    \\
\rowcolor[rgb]{0.733,0.827,0.906} \textbf{97}  & Iowa State University                         & 29            & 70           & 63           & 8            & 3            & 2.9          & 2.5              & 2.6               & 2.2                    \\
\textbf{97}                                    & Syracuse University                           & 25            & 86           & 48           & 9            & 2            & 2.8          & 2.5              & 2.5               & 2.2                    \\
\rowcolor[rgb]{0.733,0.827,0.906} \textbf{97}  & University of Delaware                        & 31            & 73           & 54           & 6            & 4            & 2.6          & 2.5              & 2.4               & 2.5                    \\
\textbf{97}                                    & University of Iowa                            & 21            & 79           & 82           & 6            & 2            & 2.8          & 2.5              & 2.6               & 2.3                    \\
\rowcolor[rgb]{0.733,0.827,0.906} \textbf{97}  & University of Massachusetts Boston            & 17            & 81           & 82           & 6            & 1            & 2.2          & 2.5              & 0                 & 2                      \\
\textbf{97}                                    & University of South Carolina                  & 28            & 65           & 73           & 10           & 1            & 2.3          & 2.5              & 2.1               & 2.2                    \\
\rowcolor[rgb]{0.733,0.827,0.906} \textbf{97}  & University of Vermont                         & 11            & 113          & 95           & 2            & 2            & 1.9          & 2.5              &                   &                        \\
\textbf{108}                                   & Brandeis University                           & 18            & 67           & 36           & 7            & 4            & 2.3          & 2.4              & 2.3               & 2.3                    \\
\rowcolor[rgb]{0.733,0.827,0.906} \textbf{108} & Brigham Young University                      & 36            & 66           & 32           & 8            & 3            & 2.4          & 2.4              & 2.2               & 2.3                    \\
\textbf{108}                                   & Clemson University                            & 38            & 62           & 51           & 8            & 3            & 2.6          & 2.4              & 2.3               & 2.2                    \\
\rowcolor[rgb]{0.733,0.827,0.906} \textbf{108} & Florida State University                      & 24            & 74           & 62           & 8            & 1            & 2.6          & 2.4              & 2.3               & 2.1                    \\
\textbf{108}                                   & Kansas State University                       & 16            & 66           & 61           & 6            & 2            & 2.3          & 2.4              & 2.2               & 1.9                    \\
\rowcolor[rgb]{0.733,0.827,0.906} \textbf{108} & University of Alabama Birmingham              & 9             & 94           & 80           & 3            & 1            & 2            & 2.4              & 0                 & 2                      \\
\textbf{108}                                   & University of North Texas                     & 33            & 61           & 63           & 8            & 2            & 1.9          & 2.4              & 0                 & 1.8                    \\
\rowcolor[rgb]{0.733,0.827,0.906} \textbf{115} & George Washington University                  & 12            & 67           & 59           & 3            & 2            & 2.7          & 2.3              & 2.3               & 2.1                    \\
\textbf{115}                                   & Louisiana State University                    & 18            & 54           & 46           & 5            & 2            & 2.1          & 2.3              & 2.1               & 2.2                    \\
\rowcolor[rgb]{0.733,0.827,0.906} \textbf{115} & University of Nevada Reno                     & 18            & 61           & 64           & 3            & 1            & 1.8          & 2.3              & 0                 & 1.9                    \\
\textbf{115}                                   & University of North Carolina Charlotte        & 19            & 53           & 55           & 4            & 2            & 2.4          & 2.3              & 2.1               & 1.9                    \\
\rowcolor[rgb]{0.733,0.827,0.906} \textbf{115} & University of Oklahoma                        & 24            & 51           & 54           & 4            & 2            & 2.2          & 2.3              & 2                 & 1.9                    \\
\textbf{115}                                   & Utah State University                         & 17            & 65           & 70           & 4            & 1            & 2            & 2.3              & 0                 & 1.8                    \\
\rowcolor[rgb]{0.733,0.827,0.906} \textbf{115} & Virginia Commonwealth University              & 23            & 58           & 50           & 7            & 1            & 2.1          & 2.3              & 0                 & 2                      \\
\textbf{115}                                   & Emory University                              & 14            & 64           & 45           & 3            & 3            & 2.7          & 2.3              &                   &                        \\
\rowcolor[rgb]{0.733,0.827,0.906} \textbf{123} & Colorado School of Mines                      & 15            & 70           & 61           & 4            & 0            & 2.6          & 2.2              & 2.1               & 2                      \\
\textbf{123}                                   & Missouri University of Science Technology     & 14            & 46           & 57           & 3            & 1            & 2.1          & 2.2              & 2                 & 2.2                    \\
\rowcolor[rgb]{0.733,0.827,0.906} \textbf{123} & University of Arkansas Fayetteville           & 22            & 59           & 51           & 4            & 1            & 1.9          & 2.2              & 0                 & 2                      \\
\textbf{123}                                   & University of Hawaii Manoa                    & 19            & 29           & 39           & 7            & 3            & 2            & 2.2              & 0                 & 2.2                    \\
\rowcolor[rgb]{0.733,0.827,0.906} \textbf{123} & University of Kansas                          & 19            & 60           & 56           & 2            & 1            & 2.5          & 2.2              & 2.3               & 2.3                    \\
\textbf{123}                                   & University of Texas San Antonio               & 20            & 54           & 52           & 3            & 1            & 2.1          & 2.2              & 0                 & 2.3                    \\
\rowcolor[rgb]{0.733,0.827,0.906} \textbf{129} & Mississippi State University                  & 16            & 48           & 38           & 2            & 1            & 1.9          & 2.1              & 0                 & 1.6                    \\
\textbf{129}                                   & Naval Postgraduate School                     & 20            & 45           & 39           & 2            & 1            & 0            & 2.1              & 2.4               & 2                      \\
\rowcolor[rgb]{0.733,0.827,0.906} \textbf{129} & Old Dominion University                       & 21            & 34           & 46           & 4            & 1            & 2            & 2.1              & 0                 & 2                      \\
\textbf{129}                                   & Oregon Health and Science University          & 7             & 66           & 76           & 1            & 0            & 1.8          & 2.1              & 2.2               & 1.8                    \\
\rowcolor[rgb]{0.733,0.827,0.906} \textbf{129} & Stevens Institute of Technology               & 20            & 54           & 53           & 3            & 0            & 2.6          & 2.1              & 2.1               & 2.1                    \\
\textbf{129}                                   & University of Missouri Kansas City            & 16            & 61           & 52           & 3            & 0            & 1.8          & 2.1              & 0                 & 1.6                    \\
\rowcolor[rgb]{0.733,0.827,0.906} \textbf{129} & University of Texas El Paso                   & 19            & 34           & 40           & 4            & 1            & 1.6          & 2.1              & 0                 & 1.9                    \\
\textbf{129}                                   & University of Tulsa                           & 15            & 37           & 53           & 3            & 1            & 1.8          & 2.1              & 0                 & 2                      \\
\rowcolor[rgb]{0.733,0.827,0.906} \textbf{129} & Western Michigan University                   & 10            & 51           & 34           & 2            & 1            & 1.5          & 2.1              & 0                 & 1.6                    \\
\textbf{138}                                   & Auburn University                             & 22            & 39           & 29           & 2            & 1            & 2.5          & 2                & 2.2               & 1.7                    \\
\rowcolor[rgb]{0.733,0.827,0.906} \textbf{138} & Claremont Graduate University                 & 5             & 66           & 43           & 2            & 0            & 1.6          & 2                & 0                 & 1.9                    \\
\textbf{138}                                   & DePaul University                             & 51            & 24           & 22           & 4            & 2            & 1.9          & 2                & 0                 & 2                      \\
\rowcolor[rgb]{0.733,0.827,0.906} \textbf{138} & Kent State University                         & 20            & 39           & 27           & 3            & 1            & 1.9          & 2                & 0                 & 1.7                    \\
\textbf{138}                                   & New Mexico State University                   & 14            & 59           & 50           & 2            & 0            & 2            & 2                & 0                 & 1.9                    \\
\rowcolor[rgb]{0.733,0.827,0.906} \textbf{138} & Texas Tech University                         & 17            & 41           & 34           & 1            & 1            & 2.1          & 2                & 0                 & 1.7                    \\
\textbf{138}                                   & University at Albany SUNY                     & 14            & 62           & 47           & 1            & 0            & 2.2          & 2                & 2.1               & 2.2                    \\
\rowcolor[rgb]{0.733,0.827,0.906} \textbf{138} & University of Alabama                         & 16            & 24           & 29           & 4            & 2            & 2.3          & 2                & 0                 & 2                      \\
\textbf{138}                                   & University of Colorado Colorado Springs       & 15            & 32           & 39           & 2            & 2            & 2            & 2                & 0                 & 1.9                    \\
\rowcolor[rgb]{0.733,0.827,0.906} \textbf{138} & University of Denver                          & 9             & 56           & 51           & 1            & 0            & 1.9          & 2                & 0                 & 1.8                    \\
\textbf{138}                                   & University of Kentucky                        & 19            & 59           & 32           & 3            & 0            & 2.3          & 2                & 2.2               & 2.2                    \\
\rowcolor[rgb]{0.733,0.827,0.906} \textbf{138} & University of Louisville                      & 18            & 32           & 25           & 4            & 2            & 1.8          & 2                & 0                 & 1.8                    \\
\textbf{138}                                   & Ohio University                               & 16            & 49           & 33           & 3            & 0            & 1.9          & 2                &                   &                        \\
\rowcolor[rgb]{0.733,0.827,0.906} \textbf{151} & University of Louisiana Lafayette             & 22            & 30           & 24           & 2            & 1            & 1.9          & 1.9              & 0                 & 1.8                    \\
\textbf{151}                                   & University of Wisconsin Milwaukee             & 12            & 57           & 44           & 1            & 0            & 2            & 1.9              & 0                 & 1.8                    \\
\rowcolor[rgb]{0.733,0.827,0.906} \textbf{153} & Florida Institute of Technology               & 27            & 19           & 18           & 1            & 1            & 1.8          & 1.8              & 0                 & 1.7                    \\
\textbf{153}                                   & Oakland University                            & 23            & 16           & 25           & 2            & 1            & 1.5          & 1.8              & 0                 & 1.9                    \\
\rowcolor[rgb]{0.733,0.827,0.906} \textbf{153} & Oklahoma State University                     & 10            & 18           & 24           & 2            & 1            & 2            & 1.8              & 0                 & 1.4                    \\
\textbf{153}                                   & Southern Methodist University                 & 8             & 49           & 20           & 2            & 0            & 2.1          & 1.8              & 2                 & 1.6                    \\
\rowcolor[rgb]{0.733,0.827,0.906} \textbf{153} & University of Cincinnati                      & 20            & 33           & 29           & 2            & 0            & 2.2          & 1.8              & 2                 & 1.8                    \\
\textbf{153}                                   & University of Maine                           & 9             & 33           & 31           & 2            & 0            & 1.8          & 1.8              & 0                 & 2                      \\
\rowcolor[rgb]{0.733,0.827,0.906} \textbf{153} & University of Mississippi                     & 7             & 33           & 32           & 2            & 0            & 1.9          & 1.8              & 0                 & 1.6                    \\
\textbf{153}                                   & Clarkson University                           & 10            & 37           & 31           & 2            & 0            & 1.7          & 1.8              &                   &                        \\
\rowcolor[rgb]{0.733,0.827,0.906} \textbf{161} & Montana State University                      & 10            & 30           & 34           & 0            & 0            & 1.7          & 1.7              & 0                 & 1.6                    \\
\textbf{161}                                   & North Dakota State University                 & 12            & 38           & 32           & 0            & 0            & 1.5          & 1.7              & 0                 & 1.5                    \\
\rowcolor[rgb]{0.733,0.827,0.906} \textbf{161} & University of Idaho                           & 14            & 33           & 31           & 0            & 0            & 1.8          & 1.7              & 0                 & 1.5                    \\
\textbf{161}                                   & University of New Orleans                     & 13            & 30           & 22           & 1            & 0            & 1.5          & 1.7              & 0                 & 1.7                    \\
\rowcolor[rgb]{0.733,0.827,0.906} \textbf{161} & Rochester Institute of Technology             & 24            & 26           & 20           & 2            & 0            & 2.7          & 1.7              &                   &                        \\
\textbf{161}                                   & San Diego State University                    & 13            & 41           & 15           & 1            & 0            & 1.7          & 1.7              &                   &                        \\
\rowcolor[rgb]{0.733,0.827,0.906} \textbf{167} & Michigan Technological University             & 22            & 22           & 28           & 0            & 0            & 2.1          & 1.6              & 0                 & 1.5                    \\
\textbf{167}                                   & New Mexico Institute of Mining and Technology & 7             & 24           & 25           & 0            & 0            & 0            & 1.6              & 0                 & 1.4                    \\
\rowcolor[rgb]{0.733,0.827,0.906} \textbf{167} & Nova Southeastern University                  & 17            & 14           & 12           & 3            & 0            & 1.3          & 1.6              & 0                 & 1.4                    \\
\textbf{167}                                   & Towson University                             & 31            & 15           & 13           & 1            & 0            & 1.6          & 1.6              & 0                 & 1.5                    \\
\rowcolor[rgb]{0.733,0.827,0.906} \textbf{167} & University of Southern Mississippi            & 12            & 10           & 13           & 1            & 1            & 1.4          & 1.6              & 0                 & 1.5                    \\
\textbf{167}                                   & University of Wyoming                         & 7             & 30           & 25           & 0            & 0            & 1.6          & 1.6              & 0                 & 1.6                    \\
\rowcolor[rgb]{0.733,0.827,0.906} \textbf{167} & Southern Illinois University                  & 12            & 27           & 28           & 0            & 0            & 1.6          & 1.6              &                   &                        \\
\textbf{167}                                   & University of New Hampshire                   & 9             & 24           & 23           & 0            & 0            & 2            & 1.6              &                   &                        \\
\rowcolor[rgb]{0.733,0.827,0.906} \textbf{167} & University of Puerto Rico Mayaguez            & 9             & 12           & 14           & 2            & 0            & 1.4          & 1.6              &                   &                        \\
\textbf{167}                                   & University of Rhode Island                    & 10            & 17           & 14           & 1            & 0            & 2            & 1.6              &                   &                        \\
\rowcolor[rgb]{0.733,0.827,0.906} \textbf{177} & Air Force Institute of Technology             & 6             & 17           & 18           & 0            & 0            & 1.8          & 1.5              & 0                 & 1.5                    \\
\textbf{177}                                   & Louisiana Tech University                     & 6             & 25           & 15           & 0            & 0            & 1.6          & 1.5              & 0                 & 1.3                    \\
\rowcolor[rgb]{0.733,0.827,0.906} \textbf{177} & University of Alabama Huntsville              & 13            & 21           & 23           & 0            & 0            & 1.8          & 1.5              & 0                 & 1.7                    \\
\textbf{177}                                   & University of Colorado Denver                 & 14            & 13           & 11           & 1            & 0            & 1.9          & 1.5              & 0                 & 1.4                    \\
\rowcolor[rgb]{0.733,0.827,0.906} \textbf{181} & University of Nebraska Omaha                  & 16            & 15           & 11           & 0            & 0            & 1.7          & 1.4              & 0                 & 1.4                    \\
\textbf{182}                                   & University of Arkansas Little Rock            & 6             & 10           & 7            & 0            & 0            & 1.7          & 1.3              & 0                 & 1.7                    \\
\rowcolor[rgb]{0.733,0.827,0.906} \textbf{183} & Bowie State University                        & 12            & 2            & 5            & 0            & 0            & 1.4          & 1.2              &                   &                        \\
\textbf{184}                                   & Indiana State University                      & 6             & 0            & 2            & 0            & 0            & 1.8          & 1.1              & 0                 & 1.1                    \\
\rowcolor[rgb]{0.733,0.827,0.906} \textbf{184} & LIU Post                                      & 3             & 0            & 1            & 0            & 0            & 1.3          & 1.1              & 0                 & 1.1  \\                           
\bottomrule
\end{longtable}
}

\end{document}